\documentstyle[12pt,aasms4] {article}
\def\etal{{\it et~al.}}
\def\ie{{\it i.e.,\ }}
\def\eg{{\it e.g.,\ }}
\def\cf{{\it cf.,\ }}

\def\kms{{km~s$^{-1}$}}
\def\kpc-1{{kpc$^{-1}$}}
\def\Mpc-1{{Mpc$^{-1}$}}
\def\s-1{{sec$^{-1}$}}
\def\pdeg2{{deg$^{-2}$}}
\def\deg{$^{\circ}$}
\def\h0{{H$_0$}}
\def\q0{{$q_0$}}
\def\z{{$z$}}

\def\Hconst#1{{$H_0 = $ {#1} {\rm km~s$^{-1}$~Mpc$^{-1}$}}}
\def\wave#1{{\rm ~$\lambda${#1}}}
\def\Halpha{{H$\alpha$}}
\def\Hbeta{{H$\beta$}}

\def\HI{{{\sc H\thinspace i}}}

\def\fOII{{{\sc [O\thinspace ii]}}}
\def\fOIII{{{\sc [O\thinspace iii]}}}
\def\I{{$I_{814}$}}
\def\V{{$V_{606}$}}
\def\Vmax{{V$_{term}$}}

\begin{document}

\title{Optical Rotation Curves of Distant Field Galaxies I : \\ 
Keck Results at Redshifts to \z\ $\sim 1^{1,2}$}

\author{Nicole P. Vogt, Duncan A. Forbes, Andrew C. Phillips, Caryl Gronwall, \\
S. M. Faber, Garth D. Illingworth, and David C. Koo}
\affil{University of California Observatories / Lick Observatory, 
Board of Studies in Astronomy and Astrophysics, 
University of California, Santa Cruz, CA 95064}

\altaffiltext{1}{Based on observations obtained at the W. M. Keck 
Observatory, which is operated jointly by the California Institute 
of Technology and the University of California.}

\altaffiltext{2}{Based in part on observations with the NASA/ESA {\it Hubble
Space Telescope}, obtained at the Space Telescope Science Institute,
which is operated by AURA, Inc., under NASA contract NAS 5--26555.}

\received{27 March 1996}
\accepted{12 April 1996}
\vskip 0.75truein
\centerline{Accepted for publication in {\it The Astrophysical Journal Letters}}

\begin{abstract}

Spatially resolved velocity profiles are presented for nine faint field
galaxies in the redshift range {0.1 $\lesssim$ \z\ $\lesssim$ 1}, based on
moderate--resolution spectroscopy obtained with the Keck~10~m telescope.  These
data were augmented with high--resolution {\it Hubble Space Telescope} images
from WFPC2, which provided V and I photometry, galaxy type, orientation, and
inclination.  The effects of seeing, slit width, and slit misalignment with
respect to galaxy major axis were modeled along with inclination for each
source, in order to derive a maximum circular velocity from the observed
rotation curve.  The lowest redshift galaxy, though highly elongated, shows a
distorted low--amplitude rotation curve that suggests a merger in progress seen
perpendicular to the collision path.  The remaining rotation curves appear
similar to those of local galaxies in both form and amplitude, implying that
some massive disks were in place at \z\ $\sim 1$.  The key result is that the
kinematics of these distant galaxies show evidence for only a {\it modest}
increase in luminosity ($\Delta$M$_B \lesssim 0.6$) compared to
velocity--luminosity (Tully--Fisher) relations for local galaxies.  

\end{abstract}


\section{Introduction}

The rotational velocity and luminosity of disk galaxies are found to be
strongly correlated (Roberts \etal~1975; Tully \& Fisher 1977).  This scaling
relation --- the ``Tully--Fisher (TF) relation'' --- provides a powerful tool
to tackle such problems as deriving \h0 (\eg Pierce \& Tully 1988), or mapping
the local galaxy streaming motions (\eg Aaronson \etal~1986).  These studies
have been confined to nearby galaxies by the use of single dish \HI\ radio
observations (\eg Haynes \& Giovanelli 1991) or optical emission line spectra
(\eg Rubin \etal~1985;  Mathewson \etal~1992).  Extending velocity width
studies to more distant galaxies would be particularly valuable to investigate
galaxy evolution (Kron 1986; van der Kruit \& Pickles 1988).  Current galaxy
evolution models range from those with mild amounts of luminosity brightening
in the past (\eg Gronwall \& Koo 1995) to those requiring more dramatic changes
(\eg Broadhurst \etal~1988, Colless \etal~1990, Glazebrook \etal~1995) to
explain the large numbers of faint blue galaxies.  By comparing a distant
sample of rotation curves to local TF relations, we can directly constrain the
global brightening of disk galaxies in the past.  

The \HI\ Tully--Fisher method is limited by the sensitivity of current radio
telescopes to \z\ $\lesssim$ 0.1.  Beyond this, two approaches have been used. 
Vogt \etal~(1993) derived rotation curves from strong optical emission lines,
for two spirals at \z~$\sim$~0.2.  Without spatial resolution, an alternative
measure of distant spiral kinematics may be extracted from the velocity widths
of emission lines.  Forbes \etal~(1996) measured velocity widths for a sample
of 18 faint field galaxies with redshifts 0.2~$<$~\z~$<$~0.84, while Colless
\etal~(1994) examined the equivalent widths of \fOII\ for a sample of 26 field
galaxies with redshifts 0.1~$\lesssim$~\z~$\lesssim$~0.7.  

This paper presents rotation curves for nine field galaxies at redshifts
{0.1~$\lesssim$~\z~$\lesssim$~1}.  This project combines spatially--resolved
spectra from the 10~m~Keck telescope with inclinations and position angles
determined from the refurbished {\it Hubble Space Telescope} (HST).  These data
provide a first glimpse into the internal kinematics of galaxies out to a
redshift of \z~$\sim$~1, or one--third to one--half (for $\Omega_0$ = 1 - 0) of
the age of the universe.  

\section{Observations}

Two galaxies were observed in 1994 September as part of a program described in
Forbes \etal~(1996).  The other seven galaxies were studied in 1995 May (see
Koo \etal~1996 for details) as part of the Deep Extragalactic Evolutionary
Probe project (DEEP;  Mould 1993; Koo 1995).  WFPC2 images are available for
both programs in the F606W (\V) and F814W (\I) filters.  Spectra were acquired
with the Low Resolution Imaging Spectrograph (LRIS; Oke \etal~1995).  The
red-blazed 600 line mm$^{-1}$ grating and a 2048$\times$2048 pixel CCD with
scales of 1.26~\AA\ and {0\farcs 215} per pixel were used.  The four brightest
galaxies were observed in long--slit mode with a {1\farcs 0} slit for 30
minutes each.  The other five were observed through a multi--object slitmask
with a slit width of {1\farcs 25} and exposures totalling 1 -- 3 hours. Seeing
was approximately {0\farcs 8} and {0\farcs 95} (FWHM) for the 1994 and 1995
observations, respectively.  The spectral range varied between 4800 -- 7400~\AA\
for the Forbes \etal\ objects, 6200 -- 8800~\AA\ for the remaining long--slit
observations, and approximately 7800~\AA\ -- 1.0~\micron\ for the slitmask
data.  Since the primary goal for all the spectroscopic observations was to
obtain redshifts, no attempt was made to align the slits with the major axis of
each galaxy.  Consequently, we must consider the effects of slit misalignment
during the analysis.

\section{Data Reduction and Analysis}

\subsection{Initial Data Reduction and Sample Selection} 

Pixel--to--pixel variations were removed by subtraction of a constant bias
level and division with flat--fields generated from dome--flat images taken
with the grating but {\it without} a slit.  The LRIS CCD suffers from fringing
at a level of up to 5\% peak--to--peak in the far red, so the spectra were also
divided by an appropriate fringing flat taken through the same slits and
grating angles as the observations.  This reduces fringing by a factor of four,
but fringing still remains a problem in the presence of strong night--sky
lines. The slitmask spectra were then corrected for slit profile variations due
to small irregularities along the slit edges, and for instrumental distortions
as determined from night--sky emission lines and the ends of slitlets. 
Wavelength calibrations were based on the stronger night--sky OH emission lines
(Vogt 1995).  The spectra were then examined for spatially--extended emission
lines.  Seven of the 41 objects in the May 1995 redshift survey (Koo
\etal~1996) had such lines, and they were augmented with the two
spatially--resolved sources from Forbes \etal~(1996).

The HST images were reduced using IRAF-based reduction and analysis techniques
(Forbes \etal~1994; Phillips \etal~1996).  Total magnitudes and colors were
measured from aperture growth curves from \V\ and \I\ data;  inclination
and position angles were estimated from the outer elliptical isophotes, and disk
scale lengths were measured along the major axis as part of bulge--to--disk
decompositions.  Error estimates are noted in Table~1, and
discussed in detail in Phillips \etal~1996.  

\subsection{Measuring the Line--of--Sight Rotation Velocity}

To improve the signal--to--noise ratio (S/N), the raw rotation curves were
smoothed with a 3--pixel (0\farcs 65) boxcar along the spatial axis before
measurement.  At each position along the slit, we then fit Gaussian profiles to
the strongest emission lines observed, ranging from \Halpha\ for the lowest-\z\
objects through \Hbeta\ and the \fOIII\wave{5007} line, to the \fOII\wave{3727}
doublet for sources with \z~$\sim$~1.  The fitting algorithm (Vogt 1995)
derives the position, amplitude, width, and a linear background for a best--fit
single Gaussian (double Gaussian for the \fOII\ doublet, with rest frame
centers fixed at 3726.2 and 3728.9~\AA), and calculates S/N for these
parameters.  Three galaxies were spatially extended in more than one emission
line, and a weighted average rotation velocity was calculated.  A velocity
profile was calculated wherever the Gaussian fit met certain minimum
requirements in profile height and width ($5 \sigma$ and $3 \sigma$);  the
typical value was $10 \sigma$ for both width and amplitude fits (see details in
Vogt 1995).  The widths were also constrained to a physically reasonable range
(1 - 10~\AA).  The faint galaxy 064$-$4412 required a relaxation of height and
width requirements to $1 \sigma$ (with typical values of $3 \sigma$) as the
\fOII\ line strength was quite weak.

\subsection{Estimating the True Rotation Velocity}

Determining the rotation curves of these distant galaxies poses some special
difficulties not encountered with nearby galaxies. The galaxies are typically
not much larger than the seeing disk ($\sim$0\farcs 8--1\farcs 0) or the
slit--width (1\farcs 0--1\farcs 25), so the resultant spectra represent a
complex combination of the spatial distribution in both velocity and emission
line surface brightness.  Moreover, the slit is usually  misaligned with the
major axis.

To interpret the data, we adopt a simple exponential disk model for each
source, where the inclination and orientation relative to the slit matches
measurements from the HST images.  The velocities in the model are assumed to
rise linearly with radius out to one disk scale length, and then to remain flat
(\cf Persic \& Salucci 1991).  The spatial distribution of the emission--line
flux is assumed to follow the disk (Kennicutt 1989), although Ryder \& Dopita
(1994) find a longer scale length than that of the continuum flux.  Thus, a
scale length of 1.5 times that measured from the HST images was used.  The
model was then convolved with an appropriate Gaussian to approximate the
seeing, masked with a model slit, and a spectral line profile calculated at
each pixel along the slit. The resulting model emission line was subjected to
the same analysis as the observed lines.  Iterative adjustments to the circular
velocity of the model were made manually until the simulated and observed
emission lines matched at the velocity extremes.  The model circular velocity,
$V_{circ}$, was then adopted as the intrinsic terminal velocity, \Vmax, of the
galaxy.  The $1 \sigma$ errors in \Vmax\ shown in Figure~\ref{TF} were
estimated by varying the inclination and position angle of each galaxy by
$\pm$~10\deg~and adopting the extrema.  Varying the distance to the elbow
radius of the modeled rotation curve by 25\% had a minimal effect upon the
value of \Vmax.  Although the model was adjusted in amplitude to fit the
velocities in the outer regions of the galaxies, in most cases it provided a
good fit at all galactocentric radii.  

\section{Results}

The data for the nine sample galaxies are listed in Table~1. 
Coordinates can be found for all galaxies in Koo \etal~(1996) and Forbes
\etal~(1966).  We assume \Hconst{75} and \q0\ = 0, though unphysical, to set
upper limits on luminosity evolution.  Restframe $BVRI$ luminosities were
derived from the \V\ -- \I\ color and redshift to select a model
(non--evolving) spectral energy distribution from the set described in Gronwall
\& Koo (1995).  The models are based on those of Bruzual \& Charlot (1993). 
Galaxies were corrected for galactic extinction, assuming extinction at \I\ is
0.45 A$_B$ and at \V\ is 0.67 A$_B$, and for internal extinction, following the
methods of Tully \& Fouqu\'e (1985) and Rubin \etal~(1985), as listed in
Table~1.  These extinction corrections were applied consistently
with the authors' prescriptions, to facilitate our final comparisons to local
TF relations.

Images of the galaxies (Figure~\ref{PLATE} ({\sc Plates XX1--2)}) show them to
be mainly normal spirals, except for 0305$-$00114 which has double nuclei, and
a few others with asymmetries.  Figure~\ref{PLATE} also shows the observed
rotation curves, the size and orientation of each slit, and the adopted model
rotation curves fit to the data.  The caption includes notes for individual
galaxies, but 074$-$2262 deserves special comment.  This low luminosity galaxy
resembles a highly--inclined, late--type barred spiral with obvious spiral
arms, bright knots of star-forming regions, and highly elongated outer
isophotes, but the lack of radial velocity variations greater than $\pm$ 25
\kms\ precludes a highly inclined disk.  The object is asymmetric, shows signs
of tidal disturbance, and has a possible perturbing companion 10\arcsec\ away
on the sky. It is also possible that this is a merger in progress between two
dwarf galaxies, observed perpendicular to the path of collision.  This galaxy
serves to illustrate that even at low redshifts, in any randomly selected
sample of galaxies there exist ``disk'' galaxies quite unsuitable for study via
the TF relation.  

\section{Discussion}

An immediate --- although not surprising --- result is that the shapes of the
rotation curves of these high redshift galaxies are similar to those of
local galaxies.  The high-redshift rotation curves are relatively symmetric,
show a ``solid-body'' rise in the inner regions, and turnover to a relatively
constant circular velocity in the outer parts. The maximum velocities (see
Table~1) are comparable to those of local spirals.  Rough
calculations yield masses between 1 and 5 $\times$ $10^{11}$ $M_{\odot}$, well
within the range of masses found for nearby spiral galaxies.  This result,
combined with the apparent disks which are obvious in the HST images, shows
conclusively that some {\it massive} disk systems were in place by \z~$\sim$ 1.

In Figure~\ref{TF} we compare the data to local TF relations in the rest--frame
$B$-band, which corresponds to \V\ at \z~$\sim$ 0.4 and to \I\ at \z~$\sim$ 0.8
(\ie the $k$ corrections are small).  The comparison in Figure~\ref{TF}$a$ is
to the TF relation (inverse fit, \ie \Vmax\ as a function of $M_B$) for 32
spiral galaxies in the Ursa Major cluster (Pierce \& Tully 1988, 1992).  This
relationship is based on \HI\ velocity width measurements (corrected for
turbulent broadening), but we have not converted our optical--line terminal
velocities to radio widths since this correction ($\lesssim$~15~\kms) is small
compared to the optical error (\eg Mathewson \etal~1992; Giovanelli
\etal~1996).  The two sources with redshift \z~$\sim$~0.2 from Vogt
\etal~(1993) are plotted for comparison; with only ground--based imaging
available, they are less well constrained in inclination.  Excluding these and
the peculiar source 074$-$2262, we compute a weighted offset of 0.55~$\pm$~0.15
mag relative to the $B$-band TF relation of Pierce \& Tully, with a dispersion
of 0.71 mag.  This dispersion matches the quadratic summation of errors (0.65)
in the logarithmic velocity widths (0.47), the rest--frame $B$ magnitudes
(0.2), and an assumed intrinsic scatter in the TF relation (0.4) (\cf Willick
\etal~1996 and references therein), indicating that our error estimates are
consistent.  An analogous calculation for the $I$-band relation yields an
offset of 0.36~$\pm$~0.18~mag.  

Figure~\ref{TF}$b$ shows the same galaxies and the TF relation (double
regression fit) for local field galaxies of Hubble types Sa, Sb, and Sc as
published in Rubin \etal~(1985), corrected to \Hconst{75}.  Since our galaxy
types (see Table~1) are evenly distributed between Sb and Sc, we
compare to a local relation midway between the Sb and Sc, and measure an offset
of 0.38~$\pm$~0.22~mag.  

Several sample selection effects and assumptions work to make these offsets
upper limits.  First,  at high redshift, our samples are biased toward
intrinsically luminous galaxies, which will affect somewhat our results with
respect to TF relations (\cf Teerikorpi 1984).  Second, we have selected
spatially-extended objects, which will bias our sample toward larger galaxies. 
Third, we have selected objects with detectable emission lines, which will bias
the sample toward galaxies with stronger than average star formation --- and
therefore higher luminosity.  Corrections for these biases would all {\it
reduce} the true offsets.  Moreover, if we have not traced the rotation curve to
sufficiently large radii, we may have underestimated the maximum velocities, and
if galaxies were less dusty at earlier epochs, we may have overcorrected for
extinction.  These errors, if present, would also {\it reduce} the true offsets.
Finally, adopting \q0\ = 0.5 instead of \q0\ = 0 would decrease the restframe
luminosities by 0.1 -- 0.4 magnitudes, and again reduce the true offset.  

The issue of a representative local TF sample for comparison is also critical. 
Key factors include the photometry passband, the observations and analysis used
to determine internal velocities, the selection biases due to catalog limits,
the distortions due to peculiar velocities, internal extinction corrections,
and the TF fitting technique (\eg forward, inverse, or double regression). 
Note, however, that our sample lies in the central region ($-19.8 < M_B <
21.7$, $180 <$ \Vmax\ $< 290$ \kms) of the range fit in the local samples,
where the differences between various fits are minimized.  We assume an error
due to these effects of 0.35 mag and combine this with the measured
uncertainties in our offsets ($\sim 0.2$) from local TF relations to arrive at
an upper limit of $\Delta$M$_B \lesssim 0.55 \pm 0.38$ mag.  

In summary, we find an offset relative to local TF relations of $\Delta$M$_B
\sim 0.6$ mag as a strong upper limit.  We do {\it not} see an obvious trend
with redshift or morphological type, but our sample is small.  Our result is
compatible with other kinematic studies of field galaxies at intermediate
redshifts, such as Rix \etal~(1996), who find a brightening of 0.44 mag for
blue field galaxies at \z~$\sim$ 0.3.  

Forbes \etal~(1996) concluded that galaxies near \z~$\sim$ 0.5 show a surface
brightness increase of 0.6 $\pm$ 0.1 mag, while Schade \etal~(1995) find a 1.2
$\pm$ 0.25 mag increase in disky galaxies at redshifts 0.5 $<$~\z~$<$ 1.2. 
Since the average bulge--to--total ratio for the Schade \etal\ sample is $\sim$
0.1, any reductions due to a significant bulge would be small.  While the
samples are not directly comparable, our data suggest somewhat less brightening
and exclude any total (versus surface) brightening by more than 1.5 magnitudes
at the 99\% CL.  The results can be reconciled if disky galaxies of similar
mass were not much brighter at \z~$\sim$ 1 than today but had slightly smaller
disk scale lengths and thus higher surface brightness.  

This Letter presents first results in the investigation of kinematics via
rotation curves in high redshift galaxies.  We plan future observations to
increase the sample size, and will constrain slits to lie within 30\deg~of the
major axes.  We will use this larger, higher--quality data set to focus on a
better understanding of sample selection biases, and will further explore a
more analogous local TF relation.  Nevertheless, the current data clearly
indicate that optical rotation curves can be measured up to \z~$\sim$ 1 and
will provide an important constraint on our understanding of the evolution of
disk galaxies.

\acknowledgments

We thank K. Wu, R. Guzm\'an, and D. Kelson for aid with the spectra, and C. 
Mihos and J.  Van Gorkom for discussions of galaxy 074$-$2262.  DEEP was
established through the Center for Particle Astrophysics.  Funding was provided
by NSF grants AST-922540, AST-9120005, and AST-8858203; NASA grants
AR-5801.01-94A, GO-2684.04-87A, and GO-2684.05-87A.  C. G.  acknowledges
funding from an NSF Graduate Fellowship.

\pagebreak
\section*{Figure Captions}

\begin{figure}[h] 
	\caption{ ({\sc Plate}) 
\I-band images and observed velocity curves for the nine sample galaxies.  For
each galaxy, the LRIS slit width and orientation are indicated on the WFPC2
image on the left.  In the velocity information on the right, points represent
the observed velocities and the solid line is the model rotation curve.  Error
bars are internal errors derived from the line-fitting technique.  The velocity
curves are measured from \Halpha\ for sources with redshift \z~$<$ 0.4 and from
\fOII\wave{3727} for \z~$>$ 0.4 ($\bullet$).  We also present \fOIII\wave{5007}
for galaxy 084$-$2833 and \Hbeta\ for sources 0305$-$00115 and 0305$-$00114
($\circ$).  Some individual notes are : (074$-$2262) see text;  (074$-$2237) we
suffer significant slit losses from slit misalignment (14\deg) due to large
spatial extent, and see an {\it apparent} slope in the outer regions of the
observed spectrum though terminal velocity is achieved well within the velocity
profile; (084$-$2833) \Halpha\ and \fOIII\ flux distribution and velocities
match well in spatial extent, though the large position angle correction makes
the estimate of the terminal velocity uncertain; (0305$-$00115) \fOII\ data
($\circ$) are of poorer quality than the \Hbeta\ data ($\bullet$), and should
be discounted; (0305$-$00114) double nucleus and strong emission line spectra
suggests a merger in progress, though there is no apparent distortion in the
velocity profile and the faint outer isophotes are undisturbed (though lack of
faint tidal features may be a surface brightness dimming effect (\cf Mihos
1995);  (104$-$4024) evident dust lane, and low (S/N) \fOII\ doublet;
(094$-$2210) luminous (M$_B$= $-21.7$) galaxy with a clear disk, high (S/N)
\fOII\ doublet.  
}
	\label{PLATE}
\end{figure}

\pagebreak
\begin{figure}[h] 
	\caption{ 
Tully--Fisher diagrams. We show our data and sources taken from Vogt
\etal~(1993) compared to ($a$) the Pierce \& Tully (1988, 1992) relationship
based on \HI\ velocity width measurements for a restricted set of cluster
spirals; and ($b$) the relationships for Sa, Sb, and Sc field galaxies found by
Rubin \etal~(1985) from \Halpha\ emission line studies.  In ($a$) the best--fit
$B$-band relation (solid line) and $3 \sigma$ limits (dashed lines) from Pierce
\& Tully are shown.  The weighted fit to our eight Tully--Fisher candidate
galaxies (dotted line;  assuming the same slope) produces an offset of
0.55~$\pm$~0.16~mag.  In ($b$) the best--fit $B$-band relations (solid lines)
for galaxies of type Sa, Sb, and Sc from Rubin \etal~and $3 \sigma$ limits
(dashed lines) for the complete set of nearby spirals are given.  The weighted
fit to our eight Tully--Fisher candidate galaxies (dotted line) relative to the
average of the Sb and Sc relations produces an offset of 0.38~$\pm$~0.22~mag. 
In each case, data have been corrected for internal extinction in a manner
consistent with the local TF study; $M_B^{rest}$ thus differs between ($a$) and
($b$).  we assume \Hconst{75} and \q0\ = 0 throughout.  In summary, we see only
a slight offset of $\Delta$M$_B$~$\lesssim$~0.6~mag toward brighter
luminosities in our data.
}
	\label{TF}
\end{figure}

\end{document}